# Scalable laser micro- and nanostructuring of mould inserts for functional injection-moulded polymer surfaces

**P. Hauschwitz**[1*], **E. Čižmárová**[2], **J. Bobek**[3], **H. Pištěková**[4], **V. Sedlařík**[4], **M. Kocáb**[1], **R. Bičišťová**[1], **M. Procházka**[1], **J. Brajer**[1], **J. Mužík**[1], **M. Smrž**[1], **M. Chyla**[1], **T. Mocek**[1]

[1] *Hilase Centre, Institute of Physics, Academy of Sciences of the Czech Republic, Za Radnici 828, Dolni Brezany 25241, Czech Republic.*

[2] *Czech Technical University in Prague, Faculty of mechanical engineering, Technická 4, Prague 6, Dejvice 160 00, Czech Republic*

[3] *Technical University of Liberec, Institute for Nanomaterials, Advanced Technologies and Innovation, Department of Industrial Technology, Liberec, Czech Republic*

[4] *Centre of Polymer Systems, University Institute, Tomas Bata University in Zlin, tr. Tomase Bati 5678,760 01 Zlin, Czech Republic*

*corresponding author: petr.hauschwitz@hilase.cz

## Abstract:

Functional polymer surfaces with tailored wettability, antibacterial and adhesion properties are increasingly demanded across medical, packaging and consumer industries. A scalable manufacturing route—laser structuring of steel mould inserts followed by injection moulding replication—has been demonstrated, but its industrial adoption has so far been limited by the low throughput of conventional single-beam laser texturing. Here we report a selective process acceleration strategy that distinguishes between two structuring regimes and applies the most appropriate technique to each. For deep microhole drilling, throughput was increased up to 20-fold by exploiting the high repetition rate (1 MHz) of an ultrashort-pulse fibre laser in single-beam mode. For nanostructuring by laser-induced periodic surface structures (LIPSS), a line-beam shaping approach using a spatial light modulator enabled a 35-fold productivity gain, reaching processing speeds exceeding 100 $cm^2$ $min^{-1}$, while maintaining sub-micrometre fidelity. Replication experiments on polypropylene (PP), PA66 and ABS confirmed successful transfer of micro- and nanostructures, with PP showing the highest fidelity. Vacuum-assisted injection moulding further improved feature height by 56–283% across materials. Functional evaluation demonstrated that all laser-textured PP surfaces increased the static water contact angle relative to the untreated reference (up to ~134°, Wenzel regime). Structured PA66 surfaces reduced bacterial retention by up to 99.8% for E. coli and ~90% for S. aureus compared to unstructured controls. Lap-shear testing showed that laser textures increased the joint shear strength of PP bonded with a non-optimised adhesive by up to 30-fold. The presented approach provides a practical, coating-free route to functional polymer components that bridges the gap between laboratory-scale laser texturing and industrial injection moulding production.

# 1 Introduction

Functional plastic components with tailored surface properties are increasingly required in sectors such as medical devices, food packaging, microfluidics and consumer products [1-3]. In many of these applications, performance is governed not only by the bulk material but critically by the surface topography and chemistry [4]. Micro- and nanoscale texturing can be used to control wettability [5,6], reduced bacterial adhesion [7,8], enhanced tribological behaviour [9] or modified optical appearance [10,11]. Among available technologies, laser surface texturing stands out as a direct, maskless and contact-free approach capable of modifying a wide range of materials with high precision and reproducibility.

For high-volume polymer production, a particularly attractive approach is to create the desired topography once on a metallic mould insert and then replicate it by standard injection moulding [12]. This strategy combines the precision of laser processing with the cost efficiency of mass production [13]. Numerous studies have shown that micro- and nanostructures generated on steel moulds can be partially transferred into thermoplastic parts, enabling superhydrophobicity, anti-icing or antibacterial performance without additional coatings [4,14]. However, the industrial adoption of this route has so far been limited mainly by the slow throughput of conventional single-beam laser texturing. Typical processing rates for micrometre-scale features are only a few $cm^2$ $min^{-1}$, which is prohibitive for large or complex mould inserts.

Recent developments in beam-shaping technologies offer a potential pathway to overcome these limitations. Diffractive optical elements (DOEs) can split a single beam into thousands of fixed beamlets, while spatial light modulators (SLMs) enable dynamic generation of programmable intensity distributions, including multi-spot patterns or extended line beams [15,16]. These approaches allow parallelisation of the ablation process and can increase throughput by one to two orders of magnitude without compromising feature resolution. For instance, DOE-based systems have demonstrated more than 40 000 simultaneous beamlets with nanostructuring speeds approaching 1900 $cm^2 min^{-1}$ [17], while SLM-based line beams enable efficient processing of non-planar geometries and improved depth-of-focus tolerance [18].

At the same time, replication studies have shown that transferring micro- and nanostructures into polymers is inherently limited by material flow behaviour, cooling dynamics and feature geometry. High-aspect-ratio or hierarchical structures are often only partially replicated, which reduces their functional performance [12]. Process adaptations, such as vacuum-assisted injection moulding, can improve filling of fine features and enhance replication fidelity [19]. These findings indicate that achieving functional polymer surfaces requires not only precise structuring of the mould, but also appropriate control of the replication process.

Despite these advances, a key challenge remains insufficiently addressed: how to combine different laser processing strategies in a way that enables both high throughput and high structural fidelity across multiple length scales. In particular, microstructures and nanostructures impose fundamentally different requirements on the laser process. High-aspect-ratio microstructures benefit primarily from increased repetition rate and efficient material removal [20], whereas nanostructures such as LIPSS require controlled energy deposition and uniform fluence distribution [21], making them more suitable for beam-shaping approaches. A systematic strategy that distinguishes between these regimes and optimally combines available acceleration techniques is still largely missing.

In this work, we address this challenge by investigating high-speed laser structuring of steel mould inserts using ultrashort-pulse laser systems in combination with advanced beam-shaping techniques. Representative micro- and nanostructures, including arrays of microholes, micropillars and laser-induced periodic surface structures (LIPSS), as well as their hierarchical combinations, are fabricated and analysed. Particular attention is given to the development of a selective processing strategy that differentiates between structuring regimes and leverages appropriate acceleration approaches for each case. The structured mould inserts are subsequently used for injection moulding of polypropylene, PA66 and ABS parts, and the replicated surfaces are evaluated in terms of morphology, wettability, bacterial adhesion and bonding performance. This integrated approach enables linking laser processing strategies with replication behaviour and resulting surface functionality.

# 2 Materials and methods

## 2.1. Moulding Process

Tool steel moulds were designed with replaceable inserts of 27 × 25 mm, enabling rapid exchange of differently structured surfaces. The moulds produced polymer plates of 30 × 30 mm with a thickness of 3 mm (Figure 1).

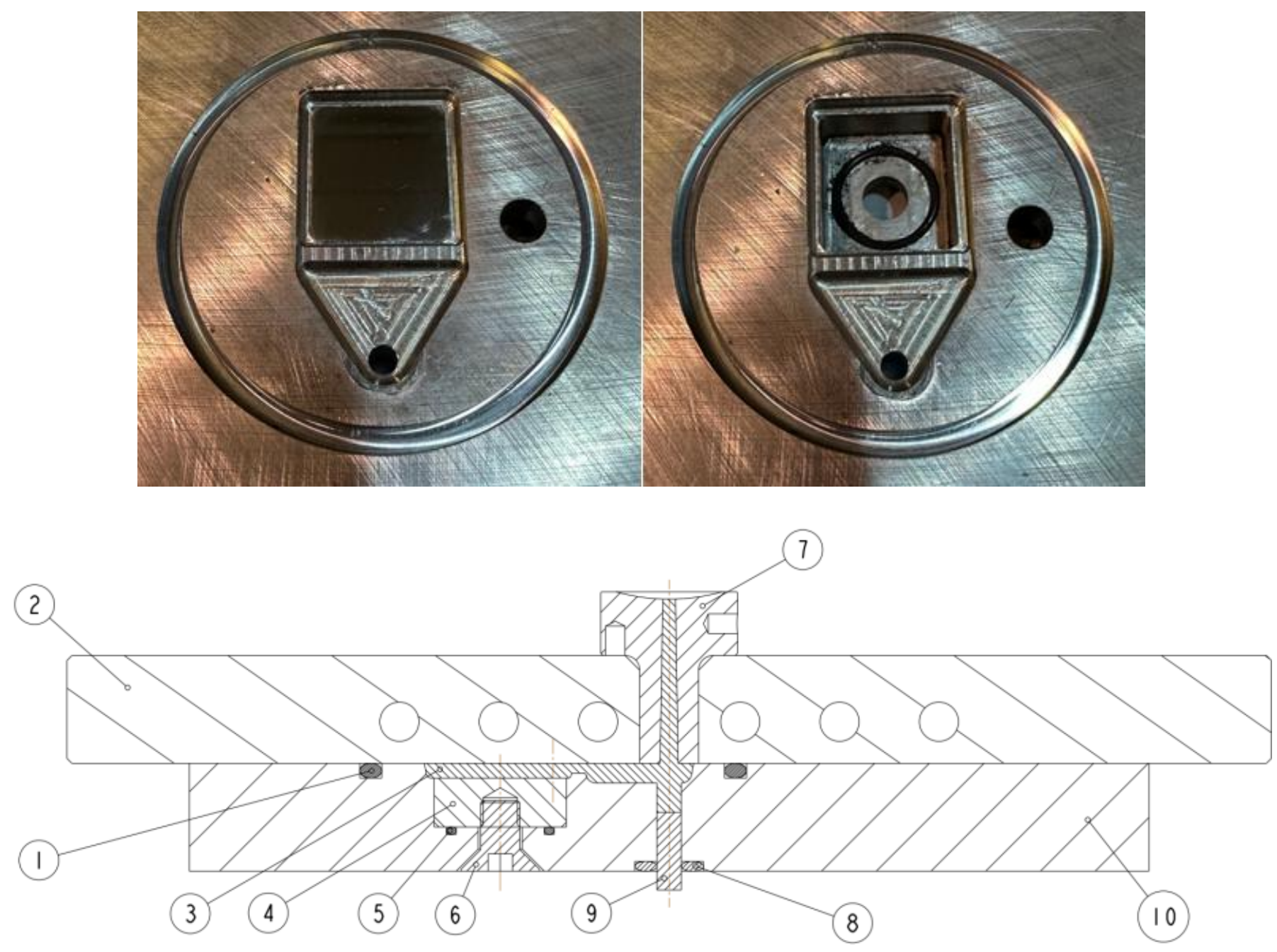

*Figure 1. Injection mould used in the study. (a) Photograph of the assembled mould showing the overall configuration. (b) Technical drawing of the mould cross-section with labelled components: 1 – split line seal; 2 – fixed mould half; 3 – injection-moulded part; 4 – mould insert with laser-modified surface; 5 – insert seal; 6 – mounting screw of insert; 7 – sprue insert; 8 – ejector seal; 9 – ejector; 10 – moving mould half.*

Injection moulding was performed on an ARBURG 270S machine equipped with a standard 25 mm screw. Three commercially available thermoplastics were selected for replication studies due to their widespread industrial use and good flow properties: PA66 (ZYTEL ST801 AHS BLACK), ABS (TERLURAN GP-22 BLACK), and PP (MOSTEN MT950 BLACK). Moulding parameters were adjusted to maximise surface replication: melt temperature at the upper limit recommended by the supplier (300 °C for PA66, 290 °C for ABS, 265 °C for PP), mould temperature 90 °C (PA66), 80 °C (ABS) and 70 °C (PP); holding pressure 80 % of the transition pressure between injection and holding phases, holding time 15 s, cooling time 20 s. All samples were demoulded with a central ejector and removed by robotic arm to avoid damaging the structured surfaces before cooling to ambient temperature

**2.2. Laser Processing of Mould Inserts**

Prior to laser processing, the inserts were cleaned with ethanol. LIPSS nanostructuring and beam-shaped processing were carried out using the ytterbium-based diode-pumped solid-state picosecond laser system Perla (HiLASE, Czech Republic), emitting 900 fs pulses at 1030 nm with $M^2$ = 1.15. The system provides a maximum average power of 100 W at a repetition rate of 50 kHz; this value describes the available system capacity rather than the process setting used for every structure.

To increase structuring efficiency, the incident beam was passed through the FBS G3 dynamic beam-shaping unit (Pulsar Photonics GmbH, Germany) equipped with a Spatial Light Modulator (SLM) from Hamamatsu Photonics, Japan. This setup allowed uploading pre-calculated computer-generated holograms (phase masks) to reshape the beam into either a line or a matrix of sub-beams. The shaped beam was directed into a galvanometer scanner (IntelliScan 14 or Scanlab GmbH, Germany) and focused onto the insert with a 100 mm telecentric F-theta lens. The resulting spot diameter on the sample was ~30 µm.

Microholes and micropillars were fabricated using an ytterbium fibre laser YLPP-100-1-100-R (IPG Photonics, USA), delivering 2 ps pulses at 1030 nm with a beam quality $M^2$ = 1.2–1.4. The laser is rated for repetition rates of 50 kHz–2 MHz, a maximum average power of 100 W and a maximum pulse energy of 100 µJ; these values define the operating envelope and are not simultaneous process settings. For the experiments reported here, the source was operated in single-beam mode at 1 MHz with pulse energies of 33–40 µJ, corresponding to average powers of 33–40 W. Beam shaping was reserved for the Perla system used to fabricate LIPSS, where sufficient pulse energy per shaped beam was available.

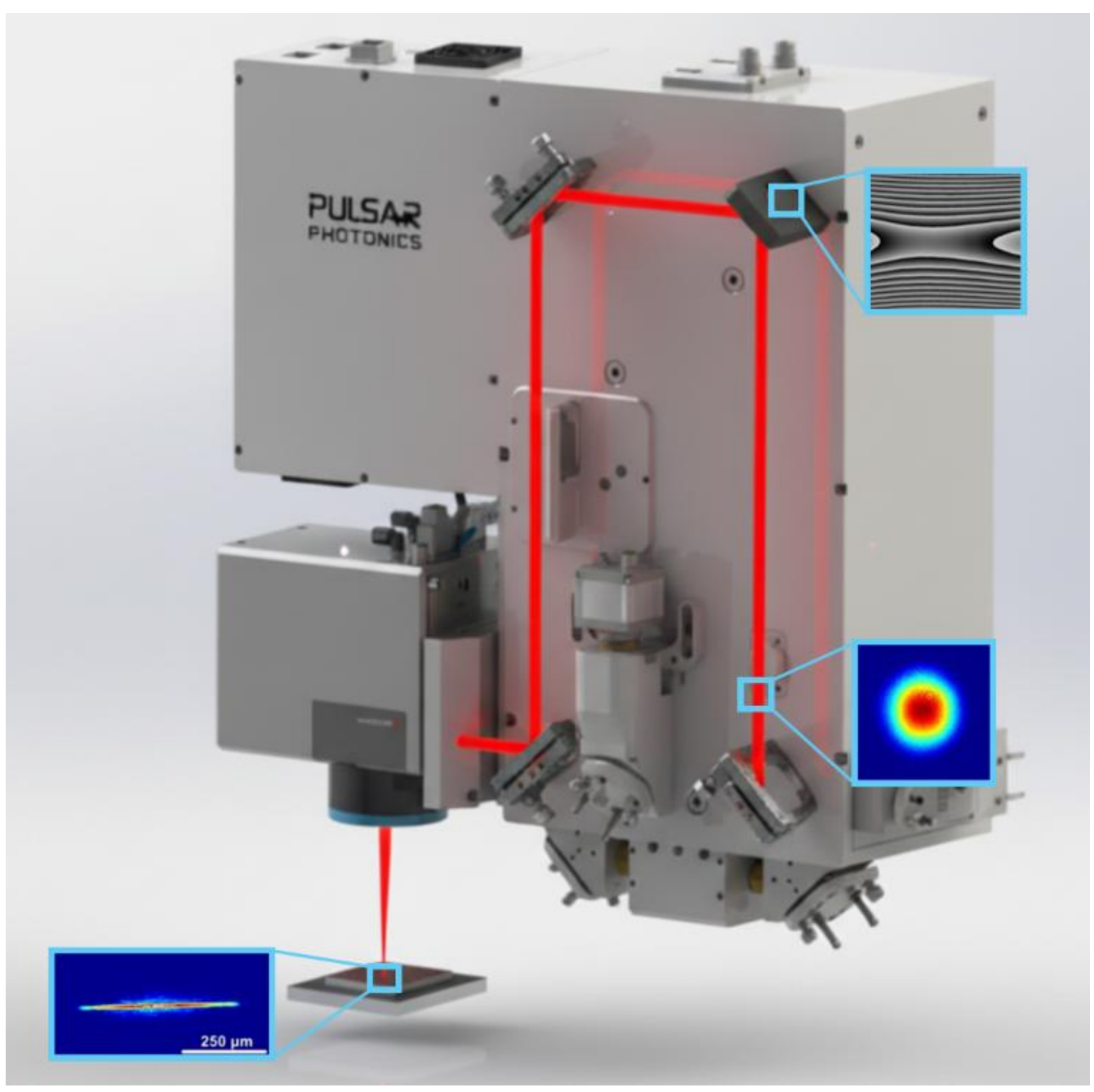

Figure 2. Schematics of the beamshaping setup with insets of input beam, computer generated hologram (CGH) and final beam at the image plane.

### 2.3. Characterisation of Produced Structures

The structured inserts and moulded polymer parts were analysed by a laser scanning confocal microscope (Olympus OLS5000) and a scanning electron microscope (Tescan MIRA, 15 keV) to evaluate feature fidelity and replication quality.

**Wettability**: Static contact angles were measured using an optical contact angle device (OCA 15EC, Data Physics Instruments) with deionised water droplets of 8 µl deposited on the sample surface.

**Antibacterial properties**: Antibacterial activity was evaluated following the procedure described by [22] with minor modifications. Cell retention tests were conducted on three identical 25×25 mm samples for each surface combination. Control samples of identical size were tested simultaneously in the same bacterial suspensions.

Prior to testing, all samples were disinfected by rinsing with 70% denatured ethanol. Two bacterial strains were used: gram-negative *Escherichia coli* CCM 4517 and gram-positive *Staphylococcus aureus* CCM 4516. Bacterial suspensions were prepared by resuspending freshly grown cultures in 500-fold diluted Nutrient Broth (NB) to an optical density of 0.5 at 562 nm (~ $1\times10^8$ CFU $mL^{-1}$; McFarland scale). Samples were submerged in 500 mL sterile containers filled with the bacterial suspension, oriented horizontally with the laser-treated surface facing upward, and incubated for 2 h at 24 °C under agitation (100 RPM, 30 mm stroke). Afterward, samples were positioned vertically for 120 sec to remove excess liquid. Residual bacteria were quantified according to ISO 18593 using orthogonal surface swabbing over a 25×25 mm area defined by sterile PTFE templates. Viable cell counts were determined by the pour plate culture method on Tryptone Soya Agar (TSA; HiMedia Laboratories, India) after incubation at 35 °C for 48 hours.

**Lap Shear Testing:** The test specimens were manufactured as single-lap joints to eliminate additional stresses caused by load eccentricity. A fixture designed in accordance with ASTM D1002 – Lap Shear Testing of Adhesively Bonded Metals was used for specimen preparation. A two-component epoxy adhesive, 3M™ Scotch-Weld™ DP190, was selected as the bonding system and applied to surfaces pre-treated by laser surface structuring. The adhesive layer thickness was controlled using glass spacer beads with a diameter of up to 200 µm.

Tensile testing was performed on a universal testing machine at a constant crosshead speed of 20 $mm\cdot min^{-1}$. The procedure followed the principles of ASTM D1002, adapted for the evaluation of joints made from non-metallic materials. During loading, the applied force and displacement were continuously recorded, and the shear strength of the joint was subsequently determined.

# 3 Results and discussion

## 3.1. Laser structuring of mould inserts

Several different surface morphologies were investigated to impart superhydrophobic and antibacterial properties on polymer surfaces. These morphologies included arrays of microholes with a diameter of 30 µm and spacing of 60 µm, square-shaped micropillars of 50 × 50 µm with a height of 30 µm, laser-induced periodic surface structures (LIPSS), and combinations of these micro- and nanostructures (Figure 3).

The microholes and micropillars were fabricated with the IPG picosecond laser in single-beam mode at 1 MHz. Microholes with a depth of approximately 100 µm required ~100 000 pulses, corresponding to ~0.10 s per hole. Micropillars were produced using pulse energies of 33–40 µJ, a scanning speed of 0.7 m $s^{-1}$, a hatch distance of 50 µm and 50 overscans. At 1 MHz, these pulse energies corresponded to average powers of 33–40 W. LIPSS were fabricated separately with the Perla laser at 50 kHz, using a fluence of 0.7 J $cm^{-2}$, a scanning speed of 1 m $s^{-1}$ and a hatch distance of 5 µm, resulting in a single-beam productivity of 3 $cm^2$ $min^{-1}$.

The productivity gain for deep microhole drilling follows directly from the repetition-rate increase. A reference process at 50 kHz requires ~2.0 s to deliver 100 000 pulses, whereas the IPG source delivered the same pulse count in 0.10 s at 1 MHz. This corresponds to a nominal 20-fold reduction in irradiation time per hole without beam splitting. The actual average power was 33 W for 33 µJ pulses and 40 W for 40 µJ pulses, remaining below the rated 100 W maximum.

For nanostructuring, the Perla laser combined with line-beam shaping produced LIPSS at 105 $cm^2$ $min^{-1}$, approximately 35× faster than the 3 $cm^2$ $min^{-1}$ single-beam process, while maintaining sub-micrometre fidelity. The accelerated processes were used for the mould structures shown in Figure 3: the IPG source produced the microholes and micropillars and the Perla line beam produced the LIPSS, including the LIPSS overlays on the hierarchical structures.

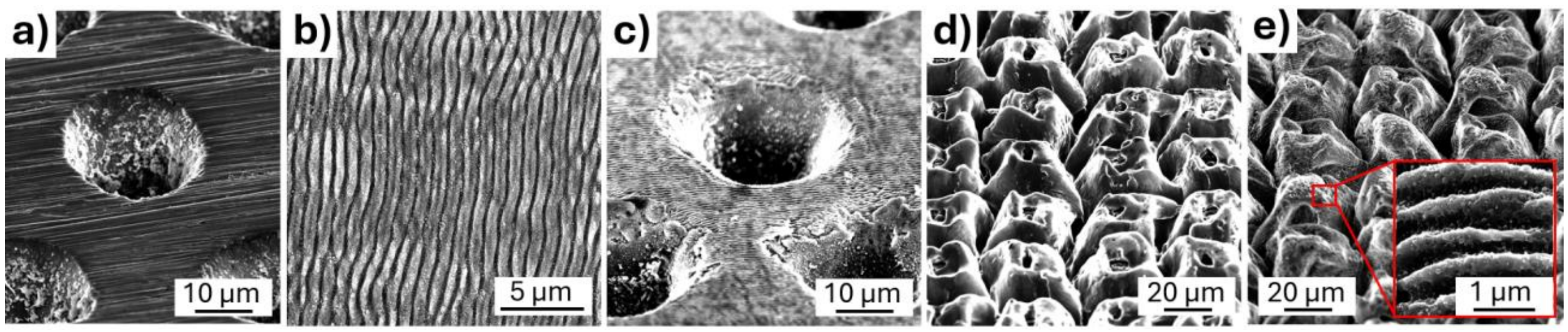

*Figure 3. Representative SEM images of optimised laser-fabricated structures on steel mould inserts produced after process acceleration. (a) Array of microholes; (b) LIPSS nanostructure; (c) combination of microholes and nanostructures; (d) array of square-shaped micropillars; (e) hierarchical micropillars covered by LIPSS*

### 3.2. Replication of micro/nanostructures into polymers

The laser-structured inserts were then mounted into the injection mould to replicate the negative of the generated structures into polypropylene, PA66 and ABS plates (30 × 30 × 3 mm).

Figure 4 shows the replicated structures in PP for all investigated morphologies, including microholes, micropillars, LIPSS and their hierarchical combinations. PP was selected as the representative material for detailed morphological analysis because it exhibited the highest replication fidelity and the most stable processing behaviour, enabling clearer visualisation of structural details. Similar trends were observed for PA66 and ABS, although with lower replication quality, as discussed below.

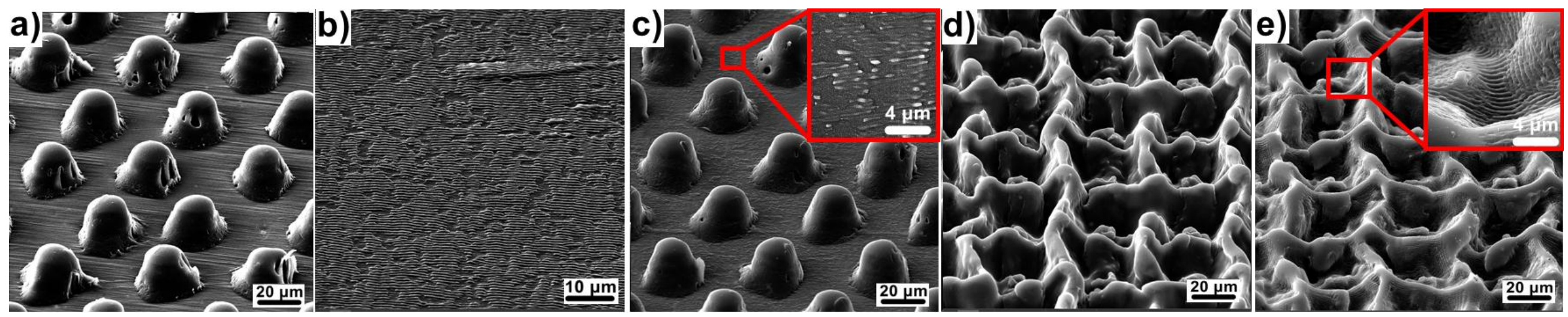

*Figure 4. Representative SEM images of replicated polypropylene surfaces for all investigated morphologies: (a) microholes, (b) LIPSS, (c) hierarchical microholes with LIPSS, (d) square-shaped micropillars, (e) hierarchical micropillars with LIPSS. PP was selected as a representative material due to its highest replication fidelity and most stable processing behaviour among the tested polymers.*

Replication quality differed between materials due to differences in melt flow and solidification behaviour. For PP, replication of microholes resulted in round micropillars with a height of ~25 µm, not reaching the full depth of the 100 µm hole. Square-shaped micropillars formed a mesh geometry whose height corresponded to the original pillars. LIPSS were clearly visible on the polymer surface, although partial loss of periodicity and local defects were observed in the hierarchical structures. Overall, PP exhibited the best LIPSS replication quality, followed by PA66 and ABS.

To overcome the limited replication of high-aspect-ratio and nanoscale features, vacuum-assisted injection moulding was implemented. This modification reduced air entrapment during filling and improved polymer flow into the finest features of the mould. The effect of this approach is illustrated in Figure 5, which compares standard and vacuum-assisted moulding for a representative hierarchical texture (microholes with LIPSS). The use of vacuum-assisted moulding led to a substantial increase in the height of replicated features: from 25 µm to 39 µm for PP (+56 %), from 13 µm to 23 µm for PA66 (+77 %), and from 3 µm to 11.5 µm for ABS (+283 %). The improvement was accompanied by a visible reduction in LIPSS defects and sharper feature definition. These observations confirm that vacuum-assisted moulding provides a general enhancement of replication fidelity across all studied materials and morphologies.

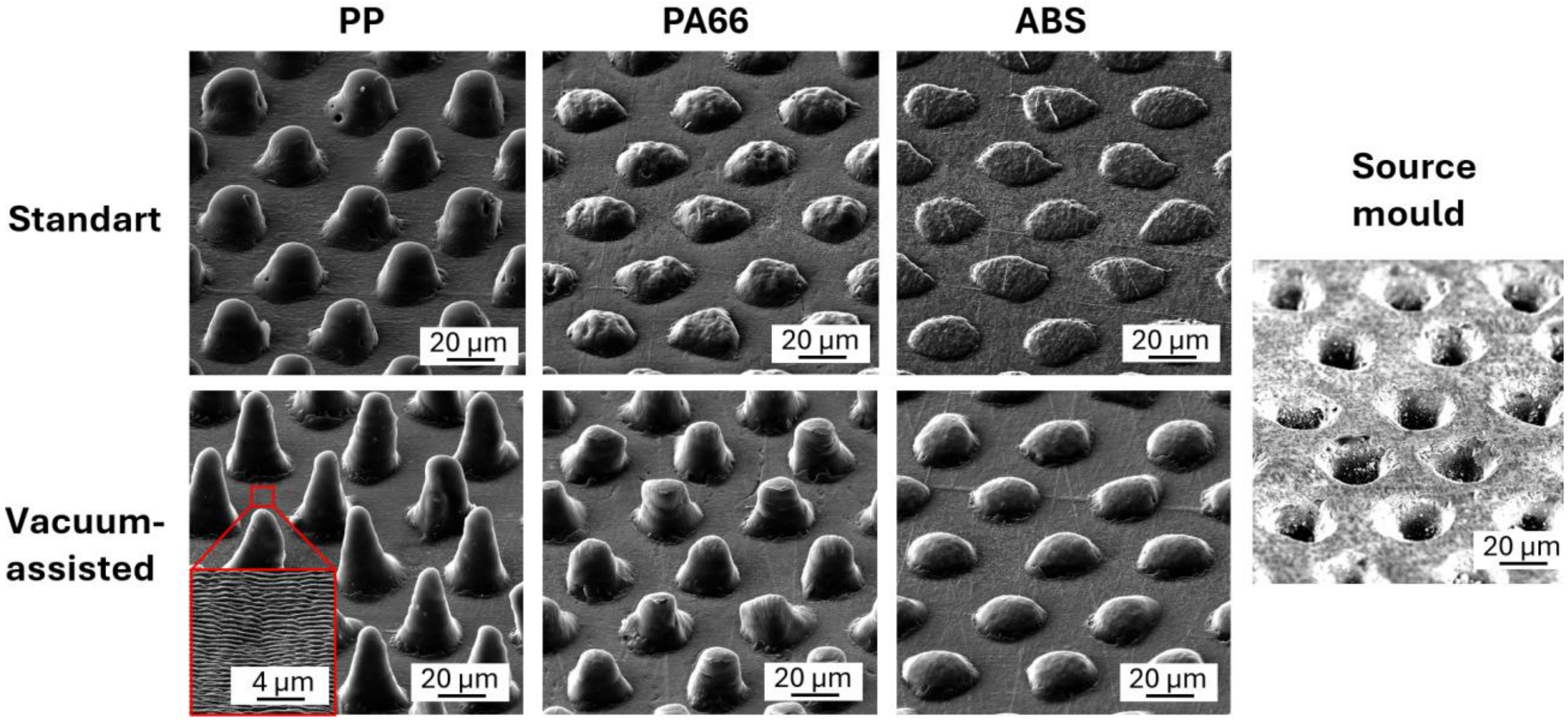

*Figure 5. Comparison between standard and vacuum-assisted injection moulding for a representative hierarchical texture (microholes with LIPSS) in different polymers. The vacuum-assisted process improved replication fidelity and feature definition, consistent with the general trend observed across all materials and morphologies.*

### 3.3. Functional properties of replicated surfaces

Functional characterisation was carried out on selected replicated polymers to evaluate the effect of laser-induced surface structures on wettability and bacterial adhesion. Wettability measurements were performed on polypropylene (PP), which exhibited the largest and most precisely replicated surface features among the tested polymers. Owing to its low surface energy and industrial importance in applications such as packaging and consumer goods, PP represents a suitable model system for assessing the influence of surface topography on apparent contact angle and wetting behaviour. Larger replicated microstructures are expected to produce more pronounced variations in wettability, allowing clearer interpretation of the observed trends.

Static contact angles measured with 8 µl deionised water droplets showed that all laser-treated surfaces increased water repellency relative to the untextured reference (108°), as shown in Figure 6. Square-mesh structures achieved the highest contact angle (133.8°). Pillar microstructures increased the contact angle by 14.5° to 122.5°. Nanostructures alone had a minor effect, increasing the contact angle by ~4° on flat surfaces and ~2° on top of microstructures. All droplets exhibited sliding angles >80°, indicating a Wenzel wetting state [23].

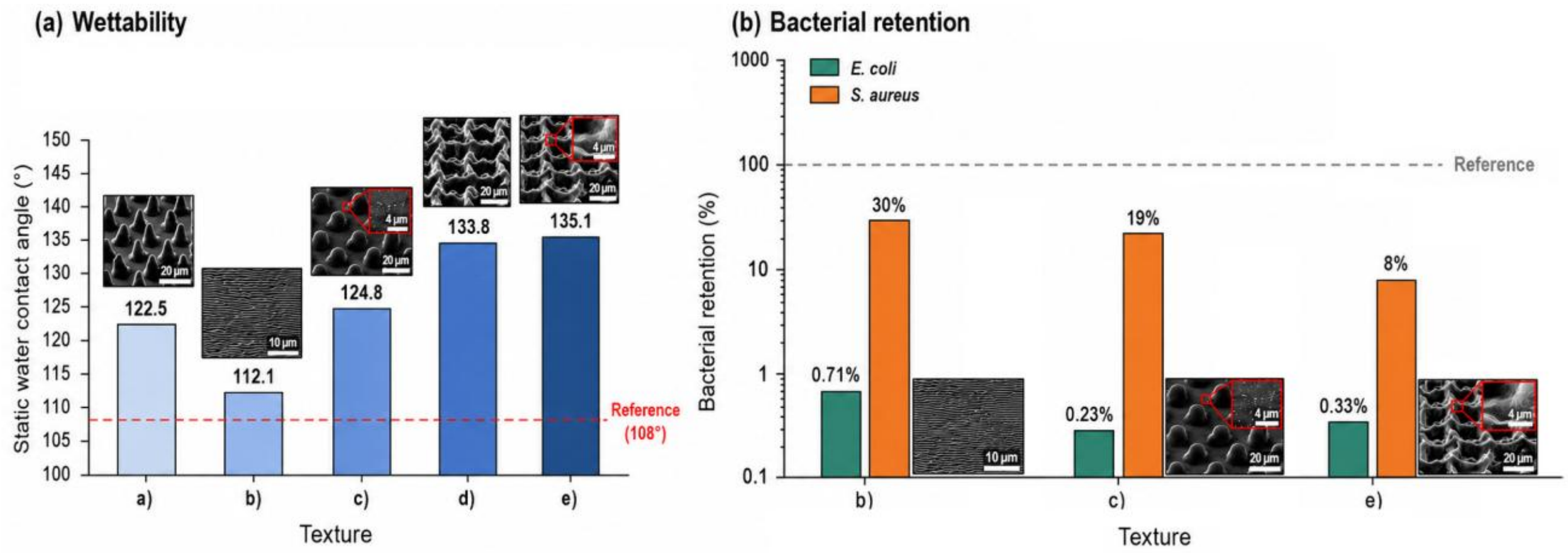

*Figure 6. Wettability and antibacterial performance of replicated surfaces with different laser-fabricated textures (a–e correspond to structures shown in Figures 3–4). **(a)** Static water contact angles measured using 8 µl deionised water droplets show enhanced hydrophobicity for all textured samples compared to the untextured reference (108°). **(b)** Bacterial retention after 2 h exposure at 24 °C to E. coli and S. aureus, normalised to the reference surface. A clear reduction in bacterial adhesion is observed for all structures, with up to 99.8 % reduction of E. coli and 92 % reduction of S. aureus.*

Antibacterial testing was conducted on polyamide 66 (PA66). This polymer combines sufficient replication fidelity with higher surface energy and moderate hydrophilicity, making it more sensitive to bacterial attachment and more relevant for evaluating antimicrobial functionality. Moreover, PA66 is widely employed in medical, automotive, and technical components where biofilm formation and hygienic performance are of practical concern. From a materials perspective, it also represents an intermediate case between PP, which yielded the largest features, and ABS, which exhibited the smallest ones, providing a balanced model for correlating topography with biological response.

Preliminary tests of bacterial adhesion were performed following the general procedure described by Lutey et al. [22]. The experiments were conducted on a limited number of samples and therefore the results should be regarded as indicative rather than conclusive. Nevertheless, all structured PA66 surfaces exhibited a clear reduction in viable bacteria compared to the untextured reference. A reduction of up to 99.8% for E. coli and approximately 90% for S. aureus was measured for textures combining micropillars and LIPSS. The lower reduction observed for S. aureus (~90%) compared to E. coli (~99.8%) is consistent with the attachment point theory described by Lutey et al. [22]. The spatial period of LIPSS (~0.7 µm) is comparable to the dimensions of both bacteria (E. coli diameter ~0.5 µm, length ~2 µm; S. aureus diameter ~0.5 µm), which geometrically limits the number of contact points available for adhesion. When surface features are of similar size to the bacterial cell, the real contact area is reduced and adhesion is weakened. In this regime, LIPSS are particularly effective compared to larger microstructures with inter-feature spacing of 20–40 µm, where bacteria can settle into the valleys and maintain contact comparable to a flat surface. The lower efficacy against S. aureus reflects its spherical morphology—a sphere rests on fewer ridge apices than a rod—and its thick, rigid Gram-positive peptidoglycan wall, which limits conformational deformation. It is also notable that wettability alone does not govern bacterial retention: Lutey et al. [22] showed that superhydrophobic spike structures performed poorly against E. coli, while LIPSS were effective even under hydrophilic conditions, confirming that sub-micron topography is the dominant driver. The anti-adhesion mechanism reported here is not bactericidal in the mechanical sense; rather, reduced contact area weakens initial attachment and limits biofilm formation. The reduction observed is consistent with the findings of Lutey et al. [22]. Our results therefore suggest that laser-induced micro/nanostructures play an important role in reducing bacterial adhesion through geometrical restriction of attachment points. Nonetheless, further systematic and statistically robust experiments are required to confirm these preliminary observations and to quantify the relative contributions of surface chemistry, wetting state and topography.

**Adhesion of bonded parts.**

To further demonstrate the functional effect of the laser-structured surfaces, lap-shear tests were performed on bonded polymer samples. The samples had surface morphology as mentioned in previous section: (a) microholes, (b) LIPSS, (c) hierarchical microholes with LIPSS, (d) square-shaped micropillars, (e) hierarchical micropillars with LIPSS. Adhesion tests were performed on acrylonitrile butadiene styrene (ABS) and polypropylene (PP) as two representative materials with markedly different surface chemistries and polarities. ABS is a polar polymer that can be effectively bonded with commercially available adhesives, whereas PP is a non-polar polymer for which suitable adhesives are limited. This contrast allowed evaluating both an optimal bonding scenario (ABS) and a non-optimal one (PP), thereby assessing how laser-generated surface textures can improve joint strength under different interfacial conditions.

When an adhesive optimised for the given material was used (ABS bonded with an ABS-specific adhesive), the shear strength remained relatively stable across the various laser-treated geometries (a–e), with sample ABS c) exhibiting the highest shear strength of 8.07 MPa, representing a slight improvement of 2.47% over the untreated reference. Conversely, sample ABS e) showed a 33% lower shear strength, suggesting that certain laser parameters may induce detrimental thermal degradation or unfavourable surface morphologies for this polymer, see Figure 7.

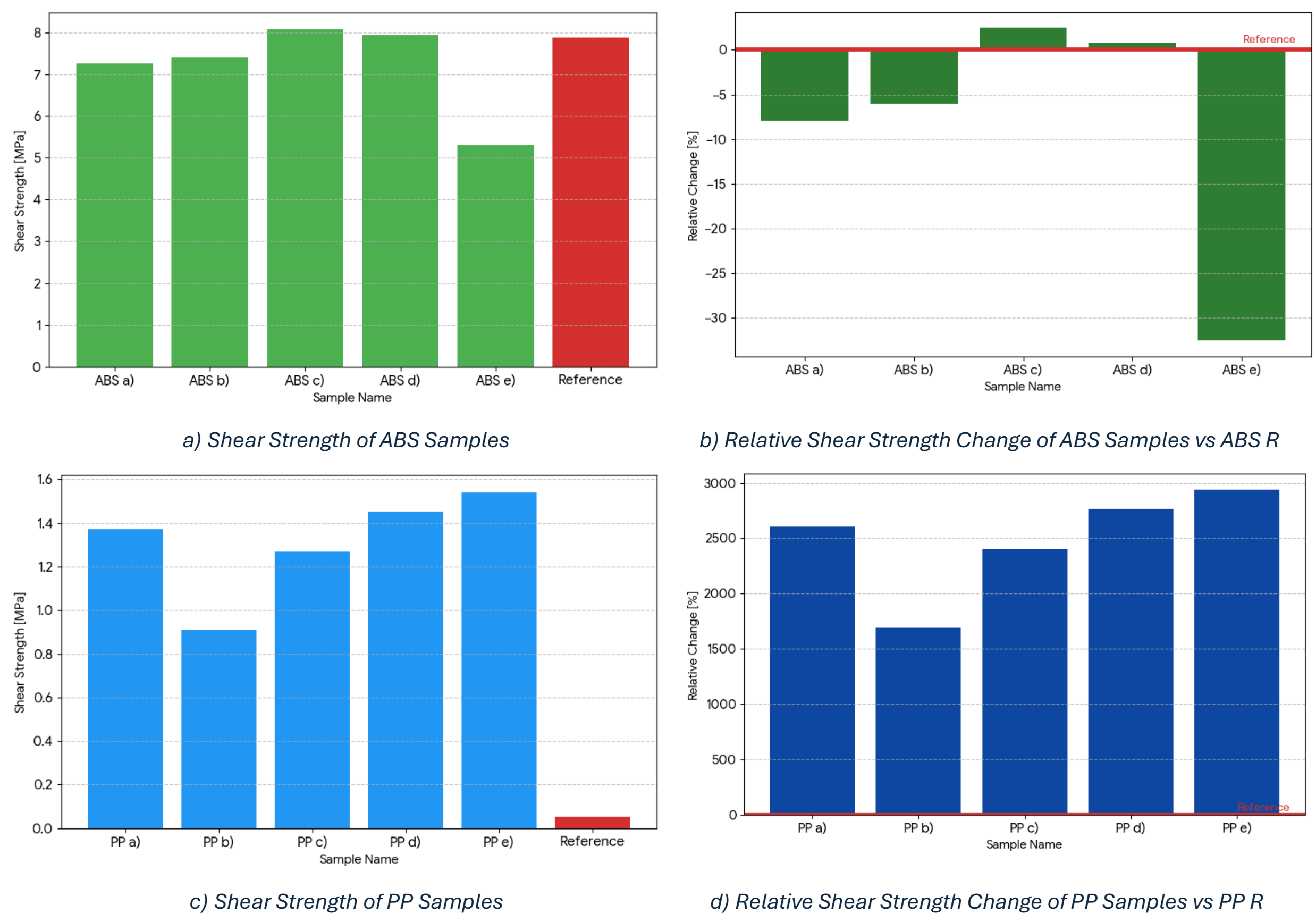

*Figure 7. Comparison of the shear strength of adhesive joints of both sample sets with the reference sample.*

In contrast, when no dedicated adhesive was available (polypropylene bonded with the same adhesive as used for ABS to simulate a non-optimal situation), the effect of the texture on joint strength was much more pronounced. In this case, shear strength improvements of up to 30-fold were recorded compared with the reference (Figure 7). These results indicate that laser-generated micro/nanostructures can improve bonding and joining performance in situations where an optimal adhesive is not available or does not exist.

The presented results confirm that high-speed laser structuring of mould inserts can be achieved with industrially relevant throughputs while preserving sub-micrometre precision and enabling precise replication of surface features into thermoplastics. Enhanced contact angles, reduced bacterial adhesion observed on the replicated polymer parts demonstrate the potential of laser-textured moulds to deliver functionalised polymer surfaces without additional coatings or chemical treatments.

A deliberate strategy for process acceleration was adopted rather than applying beam shaping indiscriminately. For the fabrication of deep microstructures, productivity was increased primarily by exploiting the higher repetition rate of a modern ultrashort-pulse source, which offers a straightforward and easily industrialised route compared with splitting the beam. In contrast, for nanostructuring such as LIPSS, where the per-feature dose is low and uniform energy distribution is critical, shaping the beam into a line provided a clear productivity advantage. By combining high-repetition-rate single-beam processing for microstructures with advanced line-beam shaping for nanostructures, an overall acceleration of up to approximately 35-fold was achieved without compromising feature quality. This selective approach illustrates how high-repetition-rate ultrashort-pulse lasers and flexible beam-shaping techniques can be combined to overcome the traditional throughput limitations of laser texturing. Together with vacuum-assisted injection moulding, which improved the replication fidelity of high-aspect-ratio and hierarchical features, the method presented here bridges the gap between laboratory-scale demonstrations and industrial implementation of functional polymer surfaces.

## Conclusion

This study demonstrates that high-precision micro- and nanostructuring of steel mould inserts for injection moulding can be achieved with industrially relevant throughput by combining ultrashort-pulse laser processing with advanced beam-shaping strategies. Well-defined microholes, micropillars and LIPSS were fabricated with sub-micrometre fidelity using picosecond laser sources.

A key outcome of this work is the identification of a selective process acceleration strategy. For high-aspect-ratio microstructures, throughput was efficiently increased by exploiting high repetition rates in single-beam mode, resulting in up to ~20× faster drilling without compromising feature quality. In contrast, for nanostructuring (LIPSS), line-beam shaping enabled a ~35× increase in productivity compared to conventional single-beam scanning, reaching processing speeds exceeding 100 $cm^2 min^{-1}$. This combination provides a practical pathway to overcome the throughput limitations that have so far hindered industrial adoption of laser-textured moulds.

Replication experiments on polypropylene, PA66 and ABS confirmed successful transfer of both micro- and nanostructures into thermoplastics, with polypropylene exhibiting the highest replication fidelity. As expected, high-aspect-ratio features were only partially replicated under standard conditions; however, vacuum-assisted injection moulding increased feature height (up to +283%) and reduced defects in hierarchical structures. These results underline that replication fidelity is not solely limited by laser processing, but can be substantially enhanced by appropriate moulding strategies.

Functional evaluation of the replicated surfaces confirmed the practical relevance of the fabricated textures. All structures increased water contact angle compared to the reference surface, with values up to ~134°, while maintaining a Wenzel wetting regime. Structured PA66 surfaces showed a strong reduction in bacterial adhesion, reaching up to 99.8% for E. coli and ~90% for S. aureus, although these results should be considered preliminary due to the limited dataset. In addition, laser-induced textures increased adhesion performance in non-ideal bonding conditions, increasing shear strength of polypropylene joints by up to 30× when using an incompatible adhesive.

Overall, the presented approach establishes a scalable route for manufacturing functional polymer surfaces by laser-textured mould inserts without the need for additional coatings or chemical treatments. By combining sub-micrometre precision with high-throughput processing and compatibility with standard injection moulding, this work bridges the gap between laboratory-scale laser structuring and industrial production of functional plastic components.

**Data Availability**

The datasets used and/or analysed during the current study are available from the corresponding author on reasonable request.

**Acknowledgements**

*This work was co-funded by the European Union and the state budget of the Czech Republic under the project LasApp CZ.02.01.01/00/22_008/0004573. Authors are further grateful for co-funding from the development process of Centre of Polymer Systems, Tomas Bata University in Zlin, program DKRVO (RP/CPS/2024-28/002) supported by the Ministry of Education Youth and Sports of the Czech Republic.*

**Author Contributions:** Conceptualization, P.H., M.K.; Methodology, P.H., E.Č., J.B. (Bobek), H.P., V.S., M.K., R.B., M.P.; Investigation, P.H., E.Č., J.B. (Bobek), H.P., V.S., M.K., R.B., M.P.; Validation, P.H., M.K., R.B., M.P.; Data curation, P.H., E.Č., J.B. (Bobek); Writing – Original Draft, P.H.; Writing – Review & Editing, P.H., M.K., R.B., M.P., J.B. (Brajer), J.M., M.S. (Smrž), M.C., T.M.; Visualization, P.H., E.Č.; Supervision, M.K., T.M.; Project administration, P.H., M.K.; Funding acquisition, T.M.

**Competing interests**

The author(s) declare no competing interests.

## References

[1] SIDDIQUIE, R. Y., A. GADDAM, A. AGRAWAL, S. S. DIMOV, et al. Anti-biofouling properties of femtosecond laser-induced submicron topographies on elastomeric surfaces. Langmuir, 2020, 36(19), 5349-5358.

[2] EVENS, T., S. CASTAGNE, D. SEVENO AND A. VAN BAEL Predicting the replication fidelity of injection molded solid polymer microneedles. International Polymer Processing, 2022, 37(3), 237-254.

[3] BASILE, V., F. MODICA, R. SURACE AND I. FASSI Micro-texturing of molds via Stereolithography for the fabrication of medical components. Procedia CIRP, 2022, 110, 93-98.

[4] PICCOLO, L., M. SORGATO, A. BATAL, S. DIMOV, et al. Functionalization of plastic parts by replication of variable pitch laser-induced periodic surface structures. Micromachines, 2020, 11(4), 429.

[5] KONG, T., G. LUO, Y. ZHAO AND Z. LIU Bioinspired Superwettability Micro/Nanoarchitectures: Fabrications and Applications. Advanced Functional Materials, 2019, 1808012.

[6] HAUSCHWITZ, P., R. JAGDHEESH, D. ROSTOHAR, J. BRAJER, et al. Hydrophilic to ultrahydrophobic transition of Al 7075 by affordable ns fiber laser and vacuum processing. Applied surface science, 2020, 505, 144523.

[7] YANG, K., J. SHI, L. WANG, Y. CHEN, et al. Bacterial anti-adhesion surface design: Surface patterning, roughness and wettability: A review. Journal of Materials Science & Technology, 2022, 99, 82-100.

[8] HAUSCHWITZ, P., Z. PALKOVÁ, L. VACHOVA, R. BICISTOVA, et al. Rapid laser-induced nanostructuring for yeast adhesion-reducing surfaces using beam shaping with SLM. Journal of Materials Research and Technology, 2025, 35, 193-198.

[9] PRIMUS, T., P. HAUSCHWITZ, T. VÍTU, R. BIČIŠŤOVÁ, et al. Enhanced tribological performance and nanostructuring speed on AlTiN by beam-shaping technology. Surface Engineering, 2022, 38(10–12), 939–947. https://doi.org/10.1080/02670844.2023.2180855.

[10] JWAD, T., P. PENCHEV, V. NASROLLAHI AND S. DIMOV Laser induced ripples' gratings with angular periodicity for fabrication of diffraction holograms. Applied surface science, 2018, 453, 449-456.

[11] FIALKOVÁ, Z., J. BRAJER, M. FLIMELOVÁ, P. HAUSCHWITZ, et al. Periodic nanogrooves from an unlikely source: nanosecond laser processing on a budget. Materials Research Express, 2025, 12(10), 105005. https://doi.org/10.1088/2053-1591/ae0d51.

[12] LUTEY, A. H., G. LAZZINI, L. GEMINI, A. PETER, et al. Insight into replication effectiveness of laser-textured micro and nanoscale morphology by injection molding. Journal of Manufacturing Processes, 2021, 65, 445-454.

[13] MASATO, D., L. PICCOLO, G. LUCCHETTA AND M. SORGATO Texturing technologies for plastics injection molding: a review. Micromachines, 2022, 13(8), 1211.

[14] GAO, P., I. MACKAY, A. GRUBER, J. KRANTZ, et al. Wetting characteristics of laser-ablated hierarchical textures replicated by micro injection molding. Micromachines, 2023, 14(4), 863.

[15] HAUSCHWITZ, P., J. MARTAN, R. BIČIŠŤOVÁ, C. BELTRAMI, et al. LIPSS-based functional surfaces produced by multi-beam nanostructuring with 2601 beams and real-time thermal processes measurement. Scientific reports, 2021, 11(1), 1-10.

[16] KUANG, Z., W. PERRIE, J. LEACH, M. SHARP, et al. High throughput diffractive multi-beam femtosecond laser processing using a spatial light modulator. Applied surface science, 2008, 255(5), 2284-2289.

[17] HAUSCHWITZ, P. High-speed multi-beam nanostructuring techniques at HiLASE with up to 40,401 beams and 1910 $cm^2$ $min^{-1}$. In Laser-based Micro- and Nanoprocessing XVI, Proceedings of SPIE, 2022, 11989, 119890O. https://doi.org/10.1117/12.2607607.

[18] MAUCLAIR, C., B. NAJIH, V. COMTE, F. BOURQUARD, et al. Dynamic spatial beam shaping for ultrafast laser processing: a review. Opto-Electronic Science, 2025, 4, 250002. https://doi.org/10.29026/oes.2025.250002.

[19] SORGATO, M., D. MASATO AND G. LUCCHETTA Effect of vacuum venting and mold wettability on the replication of micro-structured surfaces. Microsystem Technologies, 2017, 23(7), 2543-2552.

[20] ANCONA, A., S. DÖRING, C. JAUREGUI, F. RÖSER, et al. Femtosecond and picosecond laser drilling of metals at high repetition rates and average powers. Optics letters, 2009, 34(21), 3304-3306.

[21] BONSE, J. Quo vadis LIPSS?—recent and future trends on laser-induced periodic surface structures. Nanomaterials, 2020, 10(10), 1950.

[22] LUTEY, A. H., L. GEMINI, L. ROMOLI, G. LAZZINI, et al. Towards laser-textured antibacterial surfaces. Scientific reports, 2018, 8(1), 10112.

[23] WENZEL, R. N. Resistance of solid surfaces to wetting by water. Industrial & Engineering Chemistry, 1936, 28(8), 988-994.